\documentclass[twocolumn, times]{aastex63}
\usepackage{amsmath,amssymb}

\newcommand{\df}{\mathrm{d}}
\newcommand{\vct}[1]{{\boldsymbol{{#1}}}}
\newcommand{\od}[2]{\frac{\mathrm{d} {#1}}{\mathrm{d} {#2}}}
\newcommand{\pd}[2]{\frac{\partial {#1}}{\partial {#2}}}
\newcommand{\mean}[1]{\left< {#1} \right>}
\newcommand{\prt}[1]{\left( {#1} \right)}

\shorttitle{Reacceleration by downstream sound waves}
\shortauthors{Yokoyama and Ohira}

\begin{document}
\title{Particle acceleration in a shock wave propagating to an inhomogeneous medium}
\author[0000-0002-3743-414X]{Shota L. Yokoyama}
\author[0000-0002-2387-0151]{Yutaka Ohira}
\affiliation{Department of Earth and Planetary Science, The University of Tokyo, 7-3-1 Hongo, Bunkyo-ku, Tokyo 113-0033, Japan}

\received{2020 April 10}
\revised{2020 May 13}
\accepted{2020 May 15}
\submitjournal{ApJ}
\reportnum{}

\begin{abstract}
  We investigate effects of upstream density fluctuations on the diffusive shock acceleration by Monte Carlo simulations.
  The simulations show that particles are reaccelerated in the shock downstream region by a sound wave generated at the shock front.
  The time scale of turbulent acceleration by the sound wave is estimated.
  We propose a new back reaction of particles accelerated around the shock front.
  The accelerated particles generate the upstream density fluctuations by the Drury instability, which are converted to the downstream sound waves by the shock. The downstream sound waves modify the momentum spectrum of particles accelerated around the shock front.
  This new back reaction affects emission from the accelerated particles, which gives a new constraint on the acceleration efficiency of the diffusive shock acceleration.
\end{abstract}


\section{Introduction}
\label{sec1}
Cosmic rays (CRs) below the `knee' ($10^{15.5} \mathrm{eV}$) are considered to be generated in our Galaxy and propagating to the Earth.
There exists a prevailing consensus that the acceleration mechanism responsible for the production of Galactic CRs is `diffusive shock acceleration (DSA) by shock waves of supernova remnants (SNRs)'.
In DSA, particles diffusively cross the shock front many times in the turbulent magnetic field and gain energy with the momentum spectrum of ${\rm d}N/{\rm d}p\propto p^{-2}$ for the strong shock \citep{Krymsky1977, Axford1977, Bell1978, Blandford1978}.
Although many observations support DSA at SNR shocks \citep{Koyama1995, Ohira2011, Ohira2017}, there remain several problems in DSA.

\cite{Kirk2001} pointed out in their critical review of DSA three main unsolved problems of DSA ; injection problem, maximum achievable energy, and spectral index.
The latter two problems have been investigated within the context of nonlinear reactions of acceleration in most of literatures.
When the acceleration is sufficiently efficient, it is known that the shock structure is largely modified in the vicinity of shock front by the presence of accelerated particles.
Such a modified shock has a total compression ratio $r_{\rm tot}$ higher than that for the adiabatic fluid with the specific heat ratio $\gamma = 5/3$ because of the reduction of the specific heat ratio to relativistic value $\gamma = 4/3$ and the energy loss by escape of accelerated particles \citep[e.g.][]{Drury1981, Berezhko1999}.
The nonlinear reactions can produce the spectral index $s<2$ of the accelerated particles because high energy particles can diffuse far from the shock front and feel the total compression ratio $r_{\rm tot}$, while low energy particles only sense the compression of subshock, resulting in $s>2$.

It was shown that normal SNRs in the typical interstellar medium cannot accelerate particles to the energy scale of the knee \citep{Lagage1983}.
To boost the maximum energy attainable by DSA, some amplification mechanisms of magnetic field around the shock front have been proposed and studied by numerical simulations \citep{Bell2004, Malkov2010, Ohira2010, Ohira2012, Giacalone2007, Guo2012, Caprioli2013, Ohira2016, Niemiec2008, Riquelme2009, Ohira2009, Ohira2016a}.
The Bell and Drury instabilities amplify the upstream magnetic field by the accelerated particles \citep{Bell2004,Drury1986}.
The Bell instability directly amplifies the magnetic field perturbation, while the Drury instability amplifies the acoustic wave in the linear phase, resulting in the amplification of the magnetic field by the baroclinic effect in the nonlinear phase \citep{Drury2012,Downes2014}.
\citet{Malkov2010} discussed that the Drury instability grows faster than the Bell instability in the low $\beta$ plasmas, where $\beta$ is the ratio of the thermal pressure to magnetic pressure.
In any case, the accelerated particles amplify the magnetic field, so that their own maximum energy is boosted.

In addition to these nonlinear back-reactions of accelerated particles, we propose another feed back of accelerated particles in this work.
The accelerated particles amplify upstream density fluctuations by the Drury instability.
After the shock front interacts with the amplified density fluctuations, entropy, compressible and vortex modes are excited in the shock downstream region \citep{Mckenzie1968}.
The compressible and vortex modes have velocity fluctuations, that is, the turbulence is generated in the shock downstream region.
Then, the particles accelerated by DSA can be further accelerated by the turbulence.
\citet{Pohl2015} show that the second order acceleration by the wave-particle interaction in the shock downstream region can modify the energy spectrum generated by DSA at the shock front.

They mainly considered magneto-hydrodynamical waves in the downstream region whose wavelength is comparable to the mean free path of the accelerated particles.
Although they discussed acceleration by a large-scale turbulence, the momentum diffusion coefficient due to the large-scale turbulence is assumed by a model.
In this work, we consider waves longer than the particle mean free path, that are generated by the interaction between shocks and upstream density fluctuations.
In addition, we do not assume the momentum diffusion coefficient due to the large-scale turbulence, but only the particle motion in the long wave is assumed to be diffusive.

This paper is organized as follows.
In Section \ref{sec2}, we briefly review the interaction between the upstream density fluctuation and the shock wave, and acceleration processes.
We show the methods and results of our Monte Carlo simulations in Section \ref{sec3}.
In Section \ref{sec4}, the dissipation of the sound wave and other nonlinear effects are discussed.
Finally, our studies are summarized in Section \ref{sec5}.

\section{Brief review}
\label{sec2}

\subsection{Interaction between shock waves and density fluctuations}
\label{subsec21}
\cite{Mckenzie1968} analyzed a shock wave propagating in a fluctuated medium using the linearized fluid equations.
They derived amplitudes of all fluctuation modes in the downstream region for an arbitrary fluctuation mode in the upstream region.
In this work, for simplicity, we consider a monochromatic entropy mode propagating to the shock normal direction in the shock rest frame as the upstream fluctuation mode.
In this case, a sound mode and the entropy mode compressed by the shock wave are generated in the shock downstream region.
In the downstream rest frame, the sound mode propagates toward the downstream direction but the entropy mode does not propagate.
The velocity fluctuation associated with the sound mode is given by
\begin{eqnarray}
  \frac{\delta u}{u_{\rm sh}} &=& \frac{M_2}{1+2M_2 + M_1^{-2}} \cdot \frac{r-1}{r} \cdot \frac{\delta \rho_1}{\rho_1} \nonumber \\
  & \approx & \, 0.2 \, \frac{\delta \rho_1}{\rho_1} \quad ({\rm for}\, \gamma=5/3, M_1\to \infty) ,
  \label{eq01}
\end{eqnarray}
where $u_{\rm sh}, \rho_1, \delta \rho_1, M_1, M_2$ and $r=\rho_2/\rho_1$ are the shock velocity, upstream mean density, amplitude of fluctuated upstream density, upstream Mach number, downstream Mach number and compression ratio, respectively.
The entropy mode does not have any fluctuations of the velocity field.
Hereafter, the subscripts “1” and “2” represent the upstream and the downstream regions, respectively.
From the Rankine-Hugoniot relations, $r$ and $M_2$ are given by
\begin{equation}
  r = \frac{\gamma+1} {\gamma-1+ 2M_1^{-2}},
  \label{eq02}
\end{equation}
\begin{equation}
  M_2^2 = \frac{(\gamma - 1) + 2M_1^{-2}}{2\gamma - (\gamma - 1)M_1^{-2}}.
  \label{eq03}
\end{equation}
As mentioned below, the velocity fluctuation described by Equation (\ref{eq01}) is crucial for the second-order acceleration in the downstream region.
The wavelength of downstream sound mode $\lambda_{\delta u}$ is related to the upstream fluctuation,
\begin{eqnarray}
  \frac{\lambda_{\delta u_2}}{\lambda_{\delta \rho_1}} &=& \frac{1+M_2^{-1}}{r}  \nonumber \\
  &\approx& \, 0.8 \quad ({\rm for}\, \gamma=5/3, M_1\rightarrow \infty) ,
  \label{eq04}
\end{eqnarray}
where $\lambda_{\delta \rho_1}$ is the wavelength of the density fluctuation in the upstream region.

When a shock wave interacts with the upstream density fluctuation, the shock front position, $x_{\rm sh}$ is perturbed too.
This perturbation $\delta x_{\rm sh}$ is also analyzed in the linear framework of \citet{Mckenzie1968} and the relation to the upstream quantities is described as follows.
\begin{eqnarray}
  \frac{\delta x_{\rm sh}}{\lambda_{\delta \rho_1}} &=&\frac{1}{2\pi}\cdot \frac{M_2 + M_1^{-1}}{1 + 2M_2 + M_1^{-1}} \cdot \frac{\delta \rho_1}{\rho_1}   \nonumber \\
  &\approx& \, 0.04 \, \frac{\delta \rho_1}{\rho_1} \quad ({\rm for}\, \gamma=5/3, M_1\rightarrow \infty) ,
  \label{eq05}
\end{eqnarray}
It should be noted that the downstream structures described in \citet{Mckenzie1968} are not valid in the vicinity of the shock front within $\delta x_{\rm sh}$.

\subsection{Particle acceleration by shock waves}
\label{subsec22}
Propagation and acceleration of diffusive particles in a plasma flow are described by the following transport equation \citep{Parker1965}:
\begin{equation}
  \pd{f}{t} + (\vct{u} \cdot \vct{\nabla}) f-  \vct{\nabla} (\kappa \vct{\nabla} f) - \frac{1}{3} (\vct{\nabla} \cdot \vct{u}) p \pd{f}{p}= 0\ ,
  \label{eq06}
\end{equation}
where $f, \vct{u}$, and $\kappa$ are the distribution function of diffusive particles, flow velocity, and diffusion coefficient, respectively.
If the flow structure is a shock, the diffusive particles are accelerated by DSA \citep[e.g.][]{Drury1983}.
The steady-state solution of the momentum spectrum in the shock downstream region becomes
\begin{equation}
\od{N}{p}=4\pi p^2f(p) \propto p^{-s}, \quad s = \frac{r+2}{r-1}.
  \label{eq07}
\end{equation}
For $\gamma=5/3$ and the strong shock ($M_1\rightarrow \infty$), the spectral index becomes $s=2$.

\subsection{Particle acceleration by large-scale compressible turbulence}
\label{subsec23}
If the flow structure is a compressible turbulence with a length scale larger than the mean free path of the diffusive particles,
the diffusive particles are accelerated by turbulent acceleration \citep{Ptuskin1988, Bykov1982}.
In this case, the distribution function, $f$, and velocity field, $\vct{u}$, can be regarded as random variables in space.
Taking the ensemble average of Equation (\ref{eq06}) \citep[e.g.][]{Bykov1993}, we obtain
\begin{equation}
  \pd{f}{t} - \frac{1}{p^2} \pd{}{p} \prt{p^2 D_{\rm pp} \pd{f}{p}}=0,
  \label{eq08}
\end{equation}
\begin{equation}
  D_{\rm pp} = \frac{4\pi \kappa}{9} p^2 \int \df \omega \int \df k \frac{k^2 S(\omega,k)}{\omega^2 + \kappa^2 k^4}
  \label{eq09}
\end{equation}
where the spatial diffusion due to the large scale turbulence and particle diffusion are ignored and turbulence is isotropic for simplicity.
The statistical nature of the compressible turbulence is described by the spectrum, $S(\omega,k)$,
where $\omega$ and $k$ are the frequency and wavenumber of the compressible turbulence.
For the isotropic monochromatic sound wave, the momentum coefficient, $D_{\rm pp}$, can be further simplified:
\begin{equation}
  D_{\rm pp} = \frac{1}{9} p^2 \mean{\delta u^2} \frac{\kappa k^2}{v_{\rm s}^2 + (\kappa k)^2},
  \label{eq10}
\end{equation}
where $v_{\rm s}$ is the sound speed.
Note that the above equations are valid if the length and time scales that we consider are smaller than the wavelength and the period of the wave because they are derived by the ensemble average on these scales.
The momentum diffusion described by Equations (\ref{eq08}) and (\ref{eq10}) can be interpreted as the second-order Fermi acceleration.
The time-scale of second-order acceleration by compressible turbulence is given by

\begin{equation}
  t_{\rm acc,2nd} = \frac{p^2}{D_{\rm pp}}~~.
  \label{eq11}
\end{equation}

In this work, we consider particle acceleration by the large-scale downstream turbulence generated by the interaction between a shock wave and an upstream fluctuation.
From Equations (\ref{eq01}), (\ref{eq04}), and (\ref{eq10}), the acceleration time scale can be represented by
\begin{eqnarray}
  t_{\rm acc,2nd} \approx 2 & \times & 10^2 \tau_{\rm sc}(p) \cdot \prt{\frac{\delta \rho_1}{\rho_1}}^{-2} \nonumber \\
  & & \, \cdot \left\{\left(\frac{v}{u_{\rm sh}}\right)^2 + 0.07 \left(\frac{\lambda_{\delta u_2}}{\lambda_{\rm mfp}(p)}\right)^2\right\}~~,
  \label{eq12}
\end{eqnarray}
where $\gamma=5/3$ and $M_1\rightarrow \infty$ are assumed, and $\kappa=\tau_{\rm sc}(p)v^2/3$, $\tau_{\rm sc}(p)$ is the mean scattering time in the downstream region, $\lambda_{\rm mfp}(p)=\tau_{\rm sc}(p)v$ is the particle mean free path in the downstream region, $v$ is the speed of the particle.
The acceleration time of turbulent acceleration depends not only on the amplitude of fluctuation $\delta \rho_1/\rho_1$ but also on the shock velocity $u_{\rm sh}$ and the ratio of the wavelength of the sound wave to the mean free path, $\lambda_{\delta u_2}/\lambda_{\rm mfp}(p)$.
Interestingly, it almost does not depend on the wavelength of the sound wave if particles escape from one wavelength by diffusion faster than the oscillation period of the sound wave.

\section{Monte Carlo simulations}
\label{sec3}
\subsection{Methods}
\label{subsec31}
To investigate how energy spectra of particles accelerated by DSA are modified by the downstream turbulence excited by the upstream density fluctuation,
we perform test particle Monte Carlo simulations.
In our simulations, particles are scattered in the local fluid frame elastically and isotropically in the three-dimensional momentum space,
but the particles do not affect the background fluid that provides scattering bodies.
The simulation particles move in a straight line between each scattering.
The mean scattering time is given by $\tau_{\rm sc}(p) = \tau_0(p/p_0)^{\alpha}$, where $\alpha$ is a parameter to describe the momentum dependence and set to be $1$ in this work, $\tau_0$ is the mean scattering time of particles with $p=p_0$, and $p_0$ is the initial momentum.
The corresponding Lorentz factor is set to be $10$ in this work.

The number of particles with a momentum larger than $p$ significantly decreases with $p$ in this simulation.
To reduce Poisson noise in the momentum spectrum at large momenta, we employ a particle splitting method.
When the particle momentum exceeds thresholds, $10^n p_0 \, (n = 1,2, ...)$, the simulation particle is split into ten simulation particles.
The weight of the split particles is reduced by one-tenth.
After the splitting, they move individually and diffuse in a different way.

In this work, since we consider one-dimensional compressible velocity field as the first step, $\vct{u}=u(x,t)\vct{e}_{\rm x}$, we track the particle's position on only the x coordinate.
The simulation frame is the upstream rest frame.
The velocity field derived from the linear analysis is given by
\begin{equation}
  u(x,t) =
  \begin{cases}
    0 & (x < x_{\rm sh}(t)) \\
    (r^{-1} - 1) u_{\rm sh} + \delta u(x,t) & (x > x_{\rm sh}(t))
  \end{cases}
  \label{eq13}
\end{equation}
where $x_{\rm sh}(t) = -u_{\rm sh} t + \delta x_{\rm sh}(t)$ is the position of shock front.
For the downstream region ($x>x_{\rm sh}(t)$), the first part describes the shock structure and $\delta u(x,t)$ is the fluctuation of downstream velocity.

\subsection{Constraints of our simulations}
\label{subsec32}
In this work, we use the diffusion approximation to solve the particle motion and acceleration, so that we cannot correctly treat the particle motion and acceleration in a scale smaller than the particle mean free path.
Therefore, the particle mean free path has to be smaller than the wavelength of the downstream sound wave, $\lambda_{\rm mfp}(p) < \lambda_{\delta u_2}$.
In addition, there is the other constraint on the particle mean free path to keep our simulations valid.
As mentioned in Section \ref{subsec21}, the downstream velocity field (Equation~\ref{eq01}) derived from the linear analysis is not valid near the shock front within $\delta x_{\rm sh}$.
Thus, the downstream diffusion length, $L_{\rm diff} = \kappa_2/u_2$, has to be larger than the perturbation of the shock front position, $\delta x_{\rm sh}$.
Otherwise particles are accelerated by DSA in the incorrect velocity field, so that the momentum spectrum generated by DSA is incorrectly modified.
Therefore, the following condition has to be satisfied in our simulations.
\begin{equation}
  0.03 \, \prt{\frac{u_{\rm sh}}{v}} \cdot \prt{\frac{\delta \rho_1}{\rho_1}} < \frac{\lambda_{\rm mfp}(p)}{\lambda_{\delta u_2}} < 1~~,
  \label{eq14}
\end{equation}
where $\delta x_{\rm sh}$ was represented by using $\lambda_{\delta u_2}$ through Equations (\ref{eq04}) and (\ref{eq05}).
$\gamma = 5/3$ and $M_1 \to \infty$ are also assumed here.

\subsection{Results}
\label{subsec33}
\begin{figure}[t]
  \centering
  \includegraphics[width=85mm]{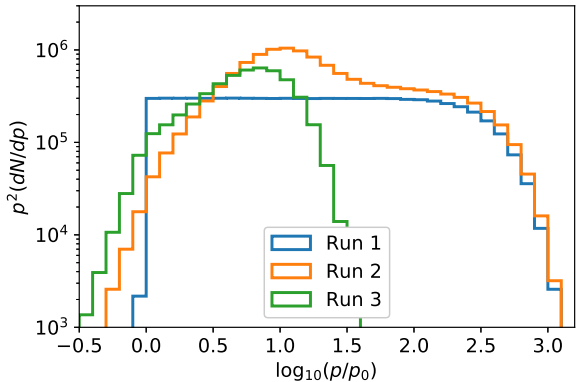}
  \caption{Momentum spectra in the whole region at $t = 2 \times 10^7 \tau_0$. The cyan and orange histograms are the spectra for the unperturbed and perturbed shocks, respectively, and for the particle injection at the shock front (Run 1 and Run 2, respectively).
  The green histogram shows the spectrum for the perturbed shock and the particle injection in the far downstream region (Run 3).}
  \label{fig1}
\end{figure}

We performed simulations for an unperturbed shock wave (Run 1) and for a shock wave with the upstream density fluctuation $\delta \rho_1 = 0.5 \, \rho_1$ (the other runs).
The simulation particles are injected to reproduce the source term of $Q(x,p,t)\propto \delta(x-x_{\rm inj})\delta(p-p_0)$, where $\delta(x)$ is the delta function.
The momentum distribution of the injected particles is isotropic and mono-momentum of $p_0$.
They are injected at constant rate at the shock front, $x_{\rm inj}=x_{\rm sh}(t)$ (Run 1 and Run 2), or at $x_{\rm inj}=10^8 \lambda_{\rm mfp}(p_0)$ (Run 3) which is too far for particles to go back to the shock front by diffusion.
The density fluctuation and the injection region for all runs are summarized in Table \ref{tab1}.
The other simulation parameters are common in all runs, $u_{\rm sh} = 0.01 \ c$ and $\lambda_{\delta u_2} = 2\times 10^2\ \lambda_{\rm mfp}(p_0)$, where $\lambda_{\rm mfp}(p) = c\tau_0(p/p_0)$ and $c$ is the speed of light.
Figure \ref{fig1} shows the momentum spectra in the whole region at $t = 2 \times 10^7 \tau_0$.
The spectrum for the unperturbed shock (Run 1, cyan histogram) is consistent with the prediction of the test particle DSA, ${\rm d}N/{\rm d}p \propto p^{-2}$.
On the other hand, for the perturbed shock (Run 2, orange histogram), the spectrum clearly deviates from ${\rm d}N/{\rm d}p \propto p^{-2}$.
Particles with $p \approx 10 \, p_0$ are accelerated more efficiently for the perturbed shock.
The green histogram shows the simulation result for the perturbed shock, where particles are injected not at the shock front,
but in the sufficiently far downstream region, $x_{\rm inj}=10^8 \lambda_{\rm mfp}(p_0)$, to avoid the DSA process (Run 3).
Even though DSA does not work, the injected particles are accelerated to $p\approx 10p_0$ in the far downstream region.
Therefore, the spectral modification observed in the simulation for the perturbed shock (orange histogram) is due to the sound wave.
As reviewed in Section \ref{subsec23}, the velocity field of the sound wave can accelerate diffusive particles by the stochastic process.
From the condition, $t_{\rm acc,2nd}=t=2 \times 10^7 \tau_0$, the typical momentum of particles accelerated by the sound wave can be estimated, which gives $p\approx 4p_0$ for our simulation conditions.
This is almost consistent with our simulation results shown in Figure \ref{fig1}.
Therefore, the mechanism of the reacceleration in the downstream region is the second-order acceleration by the sound wave.

\begin{table}[t]
  \centering
  \caption{Simulation parameters}
  \begin{tabular}{|c|c|c|}
    \hline
          & $\delta \rho_1/\rho_1$ & Injection region\\ \hline
    Run 1 & $0$                    & $x_{\rm inj}=x_{\rm sh}(t)$               \\
    Run 2 & $0.5$                  & $x_{\rm inj}=x_{\rm sh}(t)$               \\
    Run 3 & $0.5$                  & $x_{\rm inj}=10^8 \lambda_{\rm mfp}(p_0)$    \\
    Run 4 & $0.5$                  & $x_{\rm sh}(t) < x_{\rm inj} < u_2 t$     \\
    Run 5 & $0.5$                  & $x_{\rm sh}(t) < x_{\rm inj} < 0.1 u_2 t$ \\ \hline
  \end{tabular}
  \label{tab1}
\end{table}

\begin{figure}[t]
  \centering
  \includegraphics[width=85mm]{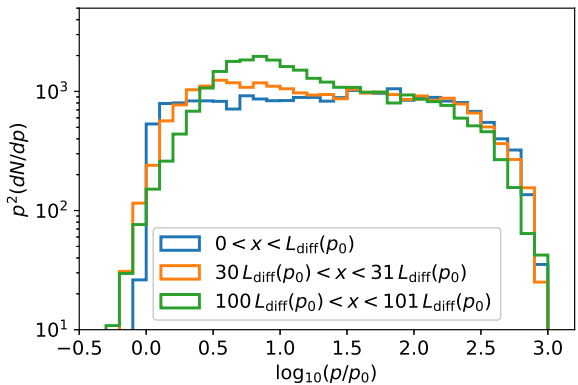}
  \caption{Momentum spectra for the perturbed shock (Run 2) in the three regions at $t = 2 \times 10^7 \tau_0$. The cyan, orange, and green histograms are the spectra in $0 < x < L_{\rm diff}(p_0)$, $30 \, L_{\rm diff}(p_0) < x < 31 \, L_{\rm diff}(p_0)$, and $100 \, L_{\rm diff}(p_0) < x < 101 \, L_{\rm diff}(p_0)$, respectively.}
  \label{fig2}
\end{figure}

\begin{figure}[t]
  \centering
  \includegraphics[width=85mm]{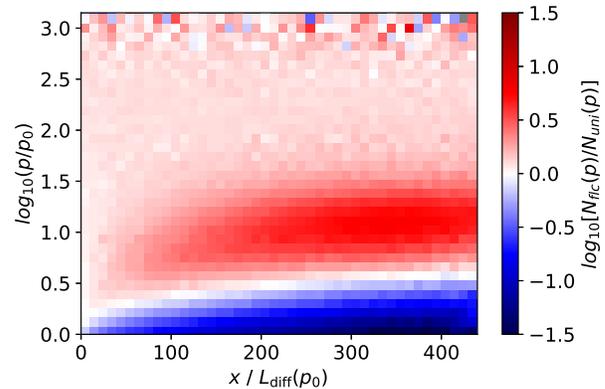}
  \caption{Ratio of the momentum spectrum for the perturbed shock (Run 2) to that for the unperturbed shock (Run 1).}
  \label{fig3}
\end{figure}

The evolution of the momentum spectrum for Run 2 in the shock downstream region is shown in Figures~\ref{fig2} and \ref{fig3}.
In Figure \ref{fig2}, the momentum spectra in three regions, $0<x<L_{\rm diff}(p_0)$(cyan), $30L_{\rm diff}(p_0)<x<31L_{\rm diff}(p_0)$(orange), $100L_{\rm diff}(p_0)<x<101L_{\rm diff}(p_0)$(green) are plotted, where $L_{\rm diff}(p_0)=\tau_0c^2/3u_2$ is the downstream diffusion length of the injected particles, $p=p_0$.
Figure \ref{fig3} shows the continuous change of the momentum spectrum in the downstream region.
The horizontal and vertical axises show the distance from the shock front and the momentum, respectively.
The color shows the ratio of the momentum spectrum for the perturbed shock (Run 2) to that for the unperturbed shock (Run 1).
As one can see, the spectral modification becomes significant as particles are advected downstream, but the spectrum does not change significantly in the vicinity of the shock front for the perturbed shock.
Therefore, our simulations show that the upstream density fluctuations do not affect the DSA process directly although the shock propagation and the downstream velocity structure become unsteady and nonuniform.
However, downstream sound waves generated by upstream density fluctuations reaccelerate particles in the shock downstream region.

\begin{figure}[t]
  \centering
  \includegraphics[width=85mm]{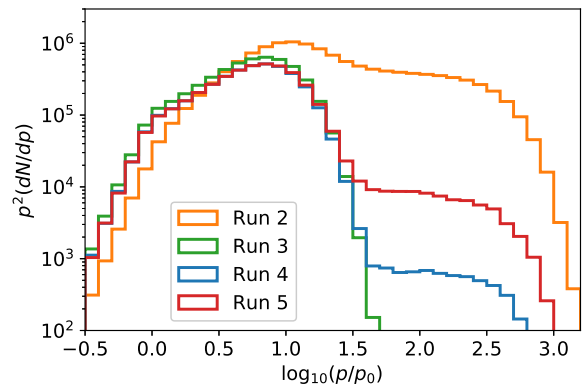}
  \caption{Momentum spectra in the whole region at $t = 2 \times 10^7 \tau_0$. Injection regions are changed as described in Table \ref{tab1}.}
  \label{fig4}
\end{figure}

Finally, we investigate effects of the spatial distribution of the particle injection.
If there is turbulence in the downstream region, particles could be accelerated by magnetic reconnection in the downstream region \citep{Zank2015}.
In addition, secondary high-energy particles such as positrons and antiprotons are injected by nuclear interaction of CR nuclei in the downstream region.
To consider such injection processes, we performed other simulations where particles are injected uniformly within the region $x_{\rm sh}(t) < x < u_2 t$ (Run 4) and $x_{\rm sh}(t) < x < 0.1 u_2 t$ (Run 5).
Simulation parameters other than the injection region are the same as those of Run 2 and Run 3.
The momentum spectra for Run 2 - Run 5 are shown in Figure \ref{fig4}.
The spectra for Run 4 and Run 5 are composed of spectrum for Run 2 and that for Run 3.
This is because particles injected in the region $x_{\rm sh}(t) < x \lesssim L_{\rm diff}(p_0)$ can go back to the shock front and be accelerated by the DSA process,
while other particles injected in the region $L_{\rm diff}(p_0) \lesssim x$ cannot reach the shock front but can be accelerated by sound waves in the downstream region.
The former and latter make the spectra same as those for Run 2 and Run 3, respectively.
The contribution of the downstream acceleration increases with the size of the downstream injection region.
Therefore, the downstream acceleration affects more on secondary CRs than on primary CRs.

\section{Discussion}
\label{sec4}
We first discuss implications of this study in CR problems.
So far, the nonlinear feed back of CRs has been thought to concave their momentum spectrum by changing the shock structure \citep{Drury1981, Berezhko1999}.
In this study, we propose another nonlinear feed back on CR momentum spectrum.
Sound waves are amplified in the CR precursor region by the Drury instability \citep{Drury1986}, so that strong turbulence is generated in the shock downstream region.
As a result, particles are accelerated by the second order acceleration by the turbulence.
Therefore, the final spectrum modified by the CR feed back becomes more complex.
Although we need more studies as discussed in the followings to understand the accurate final spectrum, we will be able to confirm the downstream acceleration by observations of SNR shocks with the high angular resolution.
If the acceleration by turbulence works in the shock downstream region, some brighting from the shock front to the downstream region should be observed.

Secondary CR acceleration by DSA in the SNR was considered to explain the hardening of secondary CRs \citep{Berezhko2003, Blasi2009, Mertsch2009, Berezhko2014}.
As shown in Figure \ref{fig4}, acceleration by the downstream turbulence affects more significantly on secondary CRs than on primary CRs.
The secondary CRs produced in the downstream region can be accelerated by the downstream turbulence efficiently compared with DSA at the shock front.
A qualitative comparison between the observed data of the secondary CRs and our study should be addressed in future.

Dissipation of the downstream sound wave has not been taken into account in this work.
In reality, however, a sound wave with a finite amplitude steepens to a shock wave, so that the sound wave is eventually dissipated by the shock dissipation.
The time scale of the shock dissipation is about $t_{\rm dis} = \lambda_{\delta u_2}/\delta u_2$ \citep{Stein1972}.
Taking the ratio of the acceleration time (Equation \ref{eq12}) to the dissipation time, we obtain
\begin{align}
  \frac{t_{\rm acc,2nd}}{t_{\rm dis}} & \approx 40 \, \prt{\frac{\delta \rho_1}{\rho_1}}^{-1} \nonumber \\
  & \cdot \left[ \prt{\frac{u_{\rm sh}}{c}}^{-1} \prt{\frac{\lambda_{\rm mfp}(p)}{\lambda_{\delta u_2}}}
  + 0.07 \, \prt{\frac{u_{\rm sh}}{c}} \prt{\frac{\lambda_{\rm mfp}(p)}{\lambda_{\delta u_2}}}^{-1} \right]~~.
  \label{eq15}
\end{align}
Note that this is a function of $(u_{\rm sh}/c)^{-1} (\lambda_{\rm mfp}(p) / \lambda_{\delta u_2})$ and has the lower limit that does not depend on the wavelength of sound wave, shock velocity and the momentum of a particle.
The lower limit of $t_{\rm acc,2nd}/t_{\rm dis}$ is given by $\sim20 (\delta \rho_1/ \rho_1)^{-1}$.
Therefore, for a one-dimensional monochromatic wave, the sound wave generated at the shock front decays more rapidly than that accelerates particles.
In contrast, for a realistic three-dimensional multi-wavelength system, in addition to the damping mechanism, sound waves with a certain wavelength is generated in the shock downstream region by the cascade process from longer waves.
Thanks to the sound waves produced by the cascade of turbulence in the far downstream region, the reacceleration by the sound waves is expected to work actually as seen in our simulations.
In addition, the shock waves formed by the steepening of sound waves, which have small Mach numbers, could reaccelerate particles by the DSA process \citep{Bykov1982, Bykov1993, Melrose1993, Inoue2010}.

For simplicity, the upstream density fluctuation is assumed to be a monochromatic wave propagating to the shock normal direction in this work.
In the real three-dimensional system, there must be obliquely propagating waves with various wavelengths in the upstream region.
In that case, incompressible vortex modes are generated in the downstream region in addition to the sound and entropy waves \citep{Mckenzie1968}.
These waves cascade from a large scale to a small scale, so that the downstream flow becomes more turbulent than that we considered in this work.
In such situations, the particle acceleration by incompressible turbulence can be expected \citep{Bykov1983, Ohira2013}.


The back reactions of the accelerated particles were not considered in this work.
It is a quite interesting and challenging problem to investigate how the downstream spectrum of particles accelerated by the nonlinear DSA is modified by the downstream turbulence when all the back reactions of the accelerated particle and nonlinear effects are considered, such as the modification of shock structure, the Drury instability, steepening of the sound waves, cascade by the downstream turbulence in the three dimensional system, and so on.
There is a possibility of rapid damping of compressible and incompressible turbulence by the back-reactions of accelerated particles \citep{Ptuskin1981, Pohl2015}.
If sufficient amount of particles are accelerated by DSA at the main shock, density fluctuations with a large amplitude are expected to be generated in the upstream region by the Drury and Bell instabilities.
Actually, recent particle simulations show the generation of the density fluctuation in the shock upstream region \citep{Caprioli2014, Bai2015, Ohira2016, Ohira2016a, VanMarle2018, VanMarle2019}.
If the spectrum of particles accelerated by DSA is significantly modified in the downstream region by turbulence excited by the upstream density fluctuation, some sort of observable signature should be expected.
Therefore, observations of SNR shocks with the high angular resolution could provide a new observational constraint on cosmic-ray acceleration efficiency.

\section{Summary}
\label{sec5}
We have proposed a new effect of back reaction of particles accelerated by DSA.
The upstream density fluctuations are generated by a back reaction of particles accelerated by DSA (e.g. Drury instability).
Then, in the shock downstream region, sound waves are generated by the interaction between the shock front and upstream density fluctuations. As a result, particles accelerated by DSA around the shock front are reaccelerated by the sound waves.
Thus, the efficient acceleration by DSA causes a modification of the momentum spectrum in the shock downstream region.
By assuming a simple downstream velocity field which is given by the linearized fluid equations, we have shown by Monte Carlo simulations that the momentum spectrum generated by DSA is significantly modified in the shock downstream region.
Furthermore, we have shown that the perturbed shock structure does not affect the DSA process directly.
Finally, we have proposed that the degree of the spectral modification in the shock downstream region provides a new constraint on the acceleration efficiency of DSA.

\acknowledgements
{The authors thank R. Yamazaki and J. Shimoda for useful comments about the interaction between the shock front and upstream density fluctuations.
Numerical computations were carried out on Cray XC50 at Center for Computational Astrophysics, National Astronomical Observatory of Japan.
SY is supported by a grant from the Hayakawa Satio Fund awarded by the Astronomical Society of Japan.
YO is supported by JSPS KAKENHI Grant Number JP16K17702 and JP19H01893, and by Leading Initiative for Excellent Young Researchers, MEXT, Japan.}

\bibliography{ref}

\end{document}